
\input harvmac.tex

\def\inbar{\,\vrule height1.5ex width.4pt depth0pt}
\def\IB{\relax{\rm I\kern-.18em B}}
\def\IC{\relax\hbox{$\inbar\kern-.3em{\rm C}$}}
\def\ID{\relax{\rm I\kern-.18em D}}
\def\IE{\relax{\rm I\kern-.18em E}}
\def\IF{\relax{\rm I\kern-.18em F}}
\def\IG{\relax\hbox{$\inbar\kern-.3em{\rm G}$}}
\def\IH{\relax{\rm I\kern-.18em H}}
\def\II{\relax{\rm I\kern-.18em I}}
\def\IK{\relax{\rm I\kern-.18em K}}
\def\IL{\relax{\rm I\kern-.18em L}}
\def\IM{\relax{\rm I\kern-.18em M}}
\def\IN{\relax{\rm I\kern-.18em N}}
\def\IO{\relax\hbox{$\inbar\kern-.3em{\rm O}$}}
\def\IP{\relax{\rm I\kern-.18em P}}
\def\IQ{\relax\hbox{$\inbar\kern-.3em{\rm Q}$}}
\def\IR{\relax{\rm I\kern-.18em R}}
\font\cmss=cmss10 \font\cmsss=cmss10 at 7pt
\def\IZ{\relax\ifmmode\mathchoice
{\hbox{\cmss Z\kern-.4em Z}}{\hbox{\cmss Z\kern-.4em Z}}
{\lower.9pt\hbox{\cmsss Z\kern-.4em Z}}
{\lower1.2pt\hbox{\cmsss Z\kern-.4em Z}}\else{\cmss Z\kern-.4em Z}\fi}
\def\IGa{\relax\hbox{${\rm I}\kern-.18em\Gamma$}}
\def\IPi{\relax\hbox{${\rm I}\kern-.18em\Pi$}}
\def\ITh{\relax\hbox{$\inbar\kern-.3em\Theta$}}
\def\IOm{\relax\hbox{$\inbar\kern-3.00pt\Omega$}}
\def\dpl{{\partial}_{+}}
\def\dm{{\partial}_{-}}
\def\xp{x^+}
\def\xm{x^-}

\Title{\vbox{\baselineskip12pt\hbox{PUPT-1340}
\hbox{hep-th/9209008}}}
{\vbox{\centerline{Semiclassical Approach}
\centerline{to}
\centerline{Black Hole Evaporation}}}

\centerline{David A. Lowe \foot{lowe@puhep1.princeton.edu}}
\centerline{\it Joseph Henry Laboratories}
\centerline{\it Princeton University}
\centerline{\it Princeton, NJ 08544}
\bigskip
\noindent
Black hole evaporation may lead to massive or
massless remnants, or naked singularities. This paper
investigates this process in the context of two quite different
two dimensional black hole models. The first is the original CGHS
model,
the second
is another two dimensional dilaton-gravity model, but with
properties much closer to physics in the real, four dimensional, world.
Numerical simulations are performed of the formation and
subsequent evaporation of  black holes and the results are found to
agree qualitatively with the exactly solved modified CGHS models,
namely that the semiclassical approximation breaks down just
before a naked singularity appears.

\Date{September, 1992}
\noblackbox

\newsec{Introduction}

Ever since Hawking's discovery
\nref\hawk{ S.W. Hawking, ``Particle creation by black holes'',
Comm. Math. Phys. {\bf 43} (1975) 199.  }%
\refs{\hawk}
of the thermal character of radiation
from black holes, much effort has been expended trying to
understand the implications of this for a theory of quantum
gravity.
Conservative suggestions have been that uncharged black holes evolve
into massive states somehow made stable by quantum gravity corrections,
or into massless remnants such as the cornucopions, in which
information is stored in a semi-infinite horn geometry. Both
these possiblities allow unitarity to be preserved
\nref\banks{T. Banks, A. Dabholkar, M.R. Douglas and
M. O'Loughlin, Phys. Rev. {\bf D45} (1992) 3607;
T. Banks and M. O'Loughlin, Rutgers preprint RU-92-14 }%
\refs{\banks}.
A radical suggestion was made by Hawking
\nref\hawk{S.W. Hawking, ``Breakdown of Predictability in
Gravitational Collapse'', Phys. Rev. {\bf D14} (1976) 2460.}%
to describe the
physics of the endpoint of black hole evaporation. It was
suggested that pure quantum states must evolve into mixed
states, leading to a breakdown in predictability.
This is an inevitable consequence of the formation of a
naked singularity when the event horizon
ceases to exist as the black hole evaporates away to leave empty
space behind. This of course violates the Cosmic Censorship
hypothesis of Penrose
\nref\penrose{R. Penrose, Rivista Nuovo Cimento, Num. Spec. I,
{\bf 1} (1969) 252.}%
\refs{\penrose}
and leaves us unable to predict the future even at the classical level
in half of spacetime.

Recently much progress has been made in developing toy models
of black hole evaporation in two dimensions. This was initiated by
the work of CGHS
\nref\cghs{C.G. Callan, S.B. Giddings, J.A. Harvey and
A. Strominger, Phys. Rev. {\bf D45} (1992) 1005.}%
\refs{\cghs}.
The quantum backreaction is included at the one loop level
by integrating out the $N$ matter fields, inducing a Polyakov-type
term
\nref\sasha{A.M. Polyakov, Phys. Lett. {\bf 103B} (1981) 211.}%
\refs{\sasha}
in the effective action.
The large $N$ limit is then taken allowing the one loop effects
of the gravitational degrees of freedom to be ignored.
Static solutions of the resulting
equations have been obtained in
\nref\hawkstat{S.W. Hawking, ``Evaporation of Two-Dimensional
Black Holes'', Phys. Rev Lett. {\bf 69} (1992) 406.}%
\nref\birnir{B. Birnir, S.B. Giddings, J.A. Harvey and
A. Strominger, ``Quantum Black Holes'',
Phys. Rev. {\bf D46} (1992) 638.}%
\nref\thor{L. Susskind and L. Thorlacius, ``Hawking Radiation and
Backreaction'', SU-ITP-92-12, hepth@xxx/9203054, March 1992.}%
\refs{\hawkstat{--} \thor}.
More recently analytic results
for dynamical shock wave solutions
have been obtained in modified models based on a Liouville type
field theory
\nref\callan{A. Bilal and C.G. Callan, ``Liouville models
of black hole evaporation'', Princeton University preprint
PUPT-1320, hepth@xxx/9205089, May 1992.}%
\nref\deal{S.P. de Alwis, ``Quantization of a theory of 2d
dilaton gravity'', University of Colorado preprint
COLO-HEP-280, hepth@xxx/9205069, May 1992;
``Black hole physics from Liouville theory'', University
of Colorado preprint COLO-HEP-284-REV,
hepth@xxx/9206020, Jun 1992;
``Quantum Black Holes in Two Dimensions'', University of
Colorado preprint COLO-HEP-288, hepth@xxx/9207095, Jul 1992.
}%
\nref\suss{J.G. Russo, L. Susskind and L. Thorlacius, ``The
endpoint of Hawking radiation'', Stanford University preprint
SU-ITP-92-17, hepth@xxx/9206070, June 1992.}%
\refs{\callan {--} \suss}.
There it was found that a naked singularity appears as the
black hole continues evaporating to negative mass.

This paper begins by reviewing the results of CGHS and then
numerical results are presented for shock wave solutions in the
original CGHS model. They are found to be in qualitative
agreement with the analytic results of the modified models,
namely that the semiclassical approximation breaks down
just before a naked singularity appears.

A model is then introduced which is obtained by the
dimensional reduction of the four dimensional Einstein action.
See
\nref\nappi{O. Lechtenfeld and C. Nappi, ``Dilaton gravity and no hair
theorem in two-dimensions'', preprint IASSNS-HEP-92-22, Mar 1992,
hepth@xxx/9204026.
}%
\refs{\nappi}
for a discussion of a general class of two dimensional
dilaton-gravity theories. This model is then coupled to matter
generally covariant only in two dimensions allowing the
backreaction to be included in the same way as the CGHS model.
The black hole solutions are of course essentially the same
as their four dimensional analogs, so it is hoped that the study
of this model will lead to greater insights into physics in the
real world. In particular the Hawking temperature is $T=1/8 \pi M$.

This model seems even more difficult than the CGHS model to treat
analytically, so once again numerical methods are used to study
both the static solutions and dynamical shock waves. The
finite mass static solutions exhibit singular horizons, reminiscent
of the solutions of the CGHS model
\refs{\hawkstat,\birnir} .
Again shock wave solutions evolve until a naked singularity
appears, at which point the approximation breaks down, and
fluctuations in the gravitational degrees of freedom
can not be neglected.
At the very least this means observers may see high energy
quantum gravity effects at the endpoint of black hole evaporation,
but the question of what really happens
is unresolved.

\newsec{Review of CGHS model}

CGHS
\refs{\cghs}
begin with the classical action for dilaton gravity in two
spacetime dimensions
\eqn\class{
S_{cl} = {1\over \pi} \int d^2\sigma \ e^{-2 \phi} \biggl[ -2\dpl \dm \rho+
4 \dpl \phi \dm \phi - \lambda^2 e^{2 \rho} \biggr] -\half \sum_{i=1}^N
\dpl f_i \dm f_i \ ,
}
where $\phi$, $\rho$ and $f$ are the dilaton, Liouville and matter
fields respectively.

Static solutions of this action representing two dimensional black
holes are
\eqn\statclass{
e^{-2 \phi}=e^{-2 \rho}= {M \over \lambda}-\lambda^2 \xp \xm .}
Here $M$ corresponds to the ADM mass of the solution. The case
$M=0$ corresponds to
a coordinate transformation of the linear dilaton vacuum solution
\eqn\vacuum{
\rho=0\qquad {\rm and} \qquad\phi= -\lambda \sigma \ ,
}
where $\sigma= \half (\sigma^+-
\sigma^-)$ and
\eqn\coords{
e^{\lambda \sigma^+}= \lambda \xp, \qquad
e^{-\lambda \sigma^-}= -\lambda \xm \ .
}
Shock waves corresponding to a black hole with mass $M$ being formed by
collapse take the form
\eqn\shockclass{
e^{-2 \phi}=e^{-2 \rho} = -\lambda^2 \xp \xm - {M\over {\lambda \xp_0}}
(\xp-\xp_0) \theta (\xp-\xp_0) \ .
}

The key point of
\refs{\cghs}
is that
in the one loop approximation quantum corrections to the classical
action may be computed simply by integrating out the fluctuations
in the matter fields. Taking the large $N$ limit, where $N$ is the
number of matter fields means the contributions of the
gravitational degrees of freedom may be ignored. The resulting action
is
\eqn\cghsaction{
S_{eff}=S_{cl}+ {1\over {\pi}} \int d^2 \sigma \
{N\over {12}} \dpl \rho \dm \rho \ .
}

The equations of motion for the case $f_i=0$ can be written in the
convenient form
\eqn\eqmotion{
\eqalign{
\dpl \dm \phi &= {{ (1-{N\over {24}} e^{2 \phi})} \over {
(1-{N\over {12}} e^{2 \phi} )}} \biggl( 2\dpl \phi \dm \phi +
{{\lambda^2}\over 2} e^{2 \rho} \biggr) \cr
\dpl \dm \rho &= {1\over {(1-{N\over {24}} e^{2\phi})}} \dpl \dm \phi \ .\cr
}
}

It should be noted \eqmotion\ become degenerate when
$\phi = \phi_{cr} = -\half ~\log (N/12) $, where in general
a curvature singularity will appear. However, one expects
the large $N$ approximation to break down before this
singularity is reached. This may be seen by examining the
matrix $K$ of the dilaton-Liouville kinetic term
$ \dpl \Phi~ K(\phi)~ \dm \Phi$ where $\Phi$ denotes the
2-vector $(\rho, \phi )$. The determinant of this matrix is
\eqn\detk{
{\rm det} K = e^{-2 \phi} ( e^{-2 \phi} - {N\over {12}} ) \ .
}
This should be $O(N^2)$ if the one loop contributions of the
dilaton and Liouville fields are to be ignored, so one sees
the large $N$ approximation fails when $\phi_{cr}-\phi \roughly{<} O(1)$.
The quantity ${N\over {12}} e^{2\phi}$ is therefore a useful measure of the
coupling of quantum gravity. When it becomes of order one, quantum
gravity effects become strong.

The following formula for the effective mass of a shock wave solution
will  be useful in the next section
\eqn\effmass{
M_{eff} = {1\over {4 \lambda}} (1- {N\over {12}} e^{2 \phi})^{{3\over 2}}
e^{-2 \phi} R \ .
}
This agrees precisely with the Bondi mass of a shock solution
along the infall line $\xp=\xp_0$, even in regions of strong
coupling. Also in the limit $\xm \to -\infty$ this agrees with the total
initial energy of the matter shock wave. After formation, i.e. for
 $\xp > \xp_0$,  $M_{eff}$
will decrease as $\xm$ increases corresponding to energy loss
as the black hole evaporates. Therefore  an
observer moving along a line of constant $\phi$ will see all the
mass of the black hole radiated away if $R \to 0$ as $\xp$ increases.

\newsec{Numerical Results For Original CGHS Model}

It may readily be seen from the form of \eqmotion\ that for fixed
$\xp$ they reduce to ordinary differential equations for the
variables $\dpl \rho$ and $\dpl \phi$. These may then be integrated
from large negative $\xm$ in towards the singularity at
${N\over {12}} e^{2 \phi} =1$. With  $\dpl \rho$ and $\dpl \phi$ now
known the solution may be evolved a step in the $\xp$ direction.
This is of course a standard method in the solution of
hyperbolic differential equations known as the
method of characteristics. Here the characteristic lines are simply
the $\xp$ and $\xm$ directions.
Details of the precise numerical method used are
given in the appendix.

The boundary conditions imposed are that $\rho$ and $\phi$ match
onto the linear dilaton vacuum along the line $\xp=\xp_0$. In order
to specify a unique solution boundary conditions must also be
imposed at past null infinity. In practice, the conditions
are that $\dpl \phi$ and $\dpl \rho$ correspond to a finite mass
classical black hole solution for $\xp>\xp_0$ and
at some large negative value of
$\xm$ where the quantum effects are always negligible.
Note that $\lambda$ may be scaled out of the problem
so in the following $\lambda=1$.

To check that the numerical algorithm was functioning correctly
the classical shock solution was evolved and the position of
the apparent horizon $\dpl \phi=0$ was plotted in comparison
to the exact value $\xm = -M$. This is shown in
\fig{\fone}{Position of the apparent horizon for a classical
shockwave solution in the original CGHS model. The dashed line is the
numerical result, the
solid line is the exact result.}
where $M=50$. Excellent
agreement is obtained.

The typical result of evolving the semiclassical
equations for a large $N$
evaporating solution is shown in
\fig{\ftwo}{Position of the apparent horizon and the
singularity for the original CGHS model. The initial mass
of the black hole is $M=50$ and $N=480$.
The apparent horizon recedes and eventually crosses the singularity.} .
The parameters
$M=50$ and $N=480$ are thought to be indicative of the
generic case.
The apparent horizon
is seen to recede as the singularity approaches and then
crosses it.
The numerical integration breaks down near the singularity so
the path of the singularity is represented by
a line of large constant $R$. This line begins to recede to larger
values of $\xm$ as $\xp$ increases signalling the appearance of
a timelike naked singularity at $\xm=\xm_s$.
The true path of the singularity lies at slightly larger
values of $\xm$ than this line and becomes tangent to $\xm=\xm_s$
at the moment when the singularity becomes naked.
After this point
both the apparent horizon and the singularity recede beyond the
region determined by propagation of the equations.
If one follows a line of constant
$\xm$ close to the line $\xm=\xm_s$ the curvature will increase
to some maximum value and then decrease. As this line approaches
$\xm_s$ the maximum value of the curvature seems to increase
without bound.

As expected the curvature on the apparent horizon
seems to increase without bound
as shown in
\fig{\fthree}{ Curvature on the apparent horizon for the original
CGHS model, $M=50$ and $N=480$.} .
In
\fig{\ffour}{Curvature along a line of constant $\phi$ for the
original CGHS model, $M=50$ and $N=480$.}
the curvature
measured by an observer
who travels along a line of constant $\phi$ is shown.
Equation \effmass\ relates this curvature to the effective mass
of the black hole. This passes through zero becoming negative
as the naked singularity continues radiating away to negative mass.

These results are qualitatively the same as the results of
\refs{\callan{--}\suss}
in that a naked singularity appears. The semiclassical equations
break down just before this singularity appears,
by which time the mass of the black hole is of order the Planck
mass.

\newsec{Another Model: Reduction of the Einstein Action From 4d to 2d}

The starting point for this model is the four dimensional
Einstein action
\eqn\einstein{
{^{(4)}S} = {1\over {2\pi}} \int d^4 \sigma \ \sqrt{-^{(4)}g }
\ {^{(4)}R} \ .
}
Restricting to spherically symmetric field configurations
allows one to write the four dimensional metric as
\eqn\fourmetric{
^{(4)} ds^2 = -e^{2 \rho} d\xp d\xm+ e^{-2 \phi} d\Omega^2 \ ,
}
where the dilaton field $\phi$ now becomes part of the
four dimensional metric.
The four dimensional Ricci curvature scalar is then
\eqn\fourricci{
^{(4)} R = 2 e^{2 \phi} + e^{-2 \rho} (8 \dpl \dm \rho+
24 \dpl \phi \dm \phi -16 \dpl \dm \phi) \ .
}
Substituting this into \einstein\ then gives the two dimensional
dilaton gravity action
\eqn\twodil{
^{(2)}S=
{1 \over {2\pi}} \int d^2 \sigma \sqrt{-^{(2)}g} \ e^{-2 \phi} \biggl(
{^{(2)}R} + 2 (\nabla \phi)^2 + 2e^{2 \phi} \biggr) \ .
}
This model has been considered at the classical level before,
for example in
\refs{\nappi} .

The static solutions of this classical action are of course the
familiar Schwarzchild solutions. These are most easily investigated
not in conformal gauge, but in the gauge
\eqn\twometric{
^{(2)} ds^2 = -h(r) dt^2+{1\over {h(r)}} dr^2 \ .
}
Solving the equations of motion in this gauge, the most general
solution up to translations in $r$ obeying $h(r) \to 1$ as
$r\to \infty$ is
\eqn\classsoln{
\eqalign{
\phi (r)&= \phi_0 - \log (r) \cr
h(r) &=1- {{2M} \over r} \ , \cr }
}
as expected. These solutions are periodic in imaginary time
with period $8\pi M$, so have Hawking temperature
$T=1/8\pi M$ as in the four dimensional case.

If one chooses to couple this classical dilaton gravity model
to matter which is generally covariant only in two dimensions
the fluctuations in the matter fields may be integrated out in the
same way as the CGHS model to give the effective action
\eqn\effaction{
\eqalign{
S_{eff}  =
 {1\over {\pi}} \int d\xp d\xm \biggl( e^{-2 \phi} & \bigl[
-2\dpl \dm \rho+ 2\dpl \phi \dm \phi \bigr] - \half e^{2 \rho} \cr
&-\half \sum_{i=1}^N \dpl f \dm f + {N\over {12}} \dpl \rho \dm \rho
\biggr)
\ , \cr}
}
where
conformal gauge has been used
\eqn\congauge{
^{(2)}ds^2 = - e^{2\rho} d\xp d\xm  \ .
}

Ideally, one would like to include matter fields generally covariant
in four dimensions. Integrating out the fluctuations of the
matter fields in four dimensions then leads to a trace anomaly of the
stress energy tensor that includes $R^2$ type terms. The
effective action must then include highly nonlocal terms and the
problem seems to become intractable. In the following
sections the effective action \effaction\ will be studied
as a two dimensional toy model of black hole evaporation with
properties much closer to four dimensional black hole physics
than the original CGHS model.

The equations of motion that follow from \effaction\ are
\eqn\motion{
\eqalign{
\dpl \dm \phi &= \dpl \phi \dm \phi + {1\over {(1-{N\over
{24}}e^{2\phi})}} \bigl( \dpl \phi \dm \phi + {1\over 4} e^{2(\rho+\phi)}
\bigr) \cr
\dpl \dm \rho &= \dpl \dm \phi - \dpl \phi \dm \phi \cr
\dpl \dm f &=0  \ .\cr }
}
In addition there are the constraint equations
\eqn\constrain{
\eqalign{
0 &= T_{++} = e^{-2\phi} \bigl( 4 \dpl \rho \dpl \phi- 2 \dpl^2 \phi
+2 (\dpl \phi)^2 \bigr) + \half (\dpl f)^2 \cr
& \qquad \qquad -{N\over {12}} \bigl( (\dpl \rho)^2 - \dpl^2 \rho+
t_{+}(\xp) \bigr) \cr
0 &= T_{--} = e^{-2\phi} \bigl( 4 \dm \rho \dm \phi- 2 \dm^2 \phi
+2 (\dm \phi)^2 \bigr) + \half (\dm f)^2 \cr
& \qquad \qquad -{N\over {12}} \bigl( (\dm \rho)^2 - \dm^2 \rho+
t_{-}(\xm) \bigr) \ , \cr }
}
where $t_{+}$ and $t_{-}$ are integration functions to be fixed
by the boundary conditions. They arise from the nonlocality
still present in the covariant form of the effective action.

As in the CGHS model,  \motion\ degenerate at a critical value
of $\phi=\phi_{cr} = -\half~ \log(N/24)$, where in general
a curvature singularity will appear. Following similar arguments
as lead to \detk\ one finds the large $N$ approximation breaks down
in this region where quantum gravity effects become strong.

\newsec{Static Solutions of the New Model}

If one is interested in finite mass static solutions it is
natural to choose as the radial coordinate
$\sigma = \half (\xp-\xm) $. In these coordinates the event horizon
will turn out to be at $\sigma = -\infty$. The equations of motion
in the static limit become
\eqn\statmotion{
\eqalign{
\phi^{\prime \prime} &= (\phi^{\prime})^2 +
{1\over {(1-{N\over {24}} e^{2 \phi} )}} \bigl( (\phi^{\prime})^2+
 e^{2(\rho+\phi)} \bigr) \cr
\rho^{\prime \prime} &= \phi^{\prime \prime}- (\phi^{\prime})^2 \ ,\cr }
}
and the constraint equations when $f_i=0$ become
\eqn\statconstrain{
e^{-2\phi} \bigl( \rho^{\prime} \phi^{\prime} - \half \phi^{\prime
\prime} + \half (\phi^{\prime})^2 \bigr) - {N\over {48}} (
(\rho^{\prime})^2- \rho^{\prime \prime} + t) = 0 \ .
}
The vacuum corresponds to the solution
\eqn\vac{
\phi = -\log (\sigma), \qquad \qquad \rho=0 \ .
}
Solving the linearized equations of motion about the vacuum
leads to infinite mass solutions when $t \not= 0$, and when $t=0$
finite mass solutions with asymptotic form
\eqn\asymp{
\phi=-\log( \sigma) + {{2M} \over {\sigma}} \log (\sigma) \ , \qquad
\qquad \rho = -{M \over {\sigma}} \ .
}

The results of integrating \statmotion\ with boundary conditions
given by \asymp\ are shown in
\fig{\ffive}{Coupling ${N\over {24}} e^{2 \phi}$ versus the radial
coordinate  for static solutions of the model
obtained from the 4d Einstein equations with
mass $M= 8,10,15$, and $N=480$.}
and
\fig{\fsix}{Curvature versus the radial coordinate for
static solutions  of the model
obtained from the 4d Einstein equations with mass
$M=8,10,15$, and $N=480$.} .
These are very
similar to the static solutions of the CGHS model
\refs{\hawkstat{--} \thor}. The coupling increases as $\sigma$
decreases before bouncing off the region of critical coupling
$\phi_{cr} = -\half ~\log (N/24)$, and then decreases to zero coupling
as $\sigma \to -\infty$. The curvature
$R=-2 e^{-2\rho} \rho^{\prime \prime}$ rises to a maximum near this
bounce region before running off to $-\infty$ as $\sigma \to -\infty$.
As $M\to 0$,  $\phi_{max} \to \phi_{cr}$, and the solution closely
approaches the vacuum outside a region close to the origin.

Beyond the bounce region as $\sigma \to -\infty$
the $\exp(2(\rho+\phi))$ term in
\statmotion\ may be neglected allowing the equations to be solved
analytically
\eqn\stuff{
\eqalign{
e^{-2\phi} &= -{N\over {12}} \rho +a \sigma +b \cr
e^{-\phi} & \sqrt{ e^{-2\phi}- {N\over {24}} } - {N\over {24}} \log
\bigl( \sqrt{ e^{-2\phi}- {N\over {24}}} + e^{-\phi} \bigr) = -a
\sigma+c \ , \cr }
}
where $a,b $ and $c$ are constants. The asymptotic form of these
solutions as $\sigma \to -\infty$ is
\eqn\asform{
\eqalign{
e^{-2 \phi} &\sim -a \sigma + {N\over {48}} \log (-a \sigma) \cr
ds^2 &\sim { {e^{48 a \sigma /N}} \over {\sqrt{-a \sigma}}} (-d \tau^2+
d \sigma^2) \ . \cr
}
}
It is clear from the form of the metric that
$\sigma = -\infty$ is an event horizon at finite proper distance
and the curvature becomes singular there.

Another interesting class of static solutions are those with
regular event horizons.
As expected these will correspond to infinite mass black holes
supported by an incoming flux of radiation.
In this case the most convenient
radial coordinate seems to be $r^2 = - \xp \xm$, which ensures
that the horizon is at $r=0$.
The equations of motion in these coordinates become
\eqn\regmotion{
\eqalign{
\phi^{\prime \prime} + {1\over r} \phi^{\prime} &=
(\phi^{\prime})^2+ {1\over {(1-{N\over {24}} e^{2\phi})}} \bigl(
(\phi^{\prime})^2 - e^{2(\rho+\phi)} \bigr) \cr
\rho^{\prime \prime}+ {1\over r} \rho^{\prime} &=
\phi^{\prime \prime} + {1\over r} \phi^{\prime}- (\phi^{\prime})^2 \ ,
\cr }
}
and the constraint equations become
\eqn\regcons{
4 \rho^{\prime} \phi^{\prime}-2 \phi^{\prime \prime} + {2\over r}
\phi^{\prime} + 2 (\phi^{\prime})^2 = {N\over {12}} e^{2 \phi} \bigl(
(\rho^{\prime})^2 - \rho^{\prime \prime} + {1\over r} \rho^{\prime}+
{{\tilde t} \over {r^2}} \bigr) \ .
}
The boundary conditions for a regular horizon are therefore
\eqn\bcs{
\phi^{\prime}(0) = 0, \qquad \rho^{\prime}(0) =0, \qquad {\tilde t} =0,
\qquad \rho(0)=0, \qquad \phi(0)=\phi_h ,
}
where $\rho(0)$ has been set to zero by a scale transformation. The
solutions are parameterized by the value of $\phi$ on the horizon.
Only when $\phi_h < \phi_{cr}$ do the solutions approach the
vacuum \vac\ as $r \to \infty$.
The coupling $e^{2 \phi}$ for a typical solution is shown in
\fig{\fseven}{The coupling ${N\over {24}} e^{2 \phi}$
versus the radial coordinate for a regular horizon
static solution  of the model
obtained from the 4d Einstein equations.} .

As $\phi_h \to \phi_{cr}$ the solutions approach a limiting form as can
be seen from
\fig{\feight}{ $e^{-\rho}$ versus the radial coordinate for a
regular horizon static solution
of the model
obtained from the 4d Einstein equations
with $\phi_h = -2.67, -1.70,-1.52$.
Here $\phi_{cr} = -1.50$.}
where $e^{-\rho}$ is plotted for various values
of $\phi_h$.
These solutions match onto the vacuum solutions of \regmotion\
as $r\to \infty$ which are
\eqn\rinfty{
\eqalign{
e^{-\phi} &= a-b \log r \cr
\rho &= \log (b/r) \ ,\cr }
}
where $a$ and $b$ are constants.
The value of $b$ for the limiting form is
$b_0\approx 6.3$.

The evaporation of a large mass black hole may be well approximated
by a succession of these static solutions with regular horizons.
As the mass is radiated away $\phi_h$ will increase towards the
critical value $\phi_{cr}$. If one expects no naked singularity to
appear then this limiting form should approach zero temperature.
In fact, this is not true, as may be seen by a calculation of the
Hawking flux.
In order that the future horizon be regular one must demand that
$t_-(\xm)=0$. This is analogous to the condition ${\tilde t}=0$
for the static regular horizon solution.
Transforming \rinfty\ to asymptotically Minkowskian
coordinates $\sigma^+$ and $\sigma^-$ via
\eqn\coords{
\xp= e^{\sigma^+ /b_0}, \qquad \qquad \xm = -e^{-\sigma^- /b_0} \ ,
}
leads to
\eqn\stress{
\eqalign{
t_-(\sigma^-) &={N\over {24}} (t_-(\xm)+ D^{S}_{\sigma^-} (\xm) )
\biggl({{\partial \sigma^-}\over {\partial x^-}}\biggr)^{-2} \cr
            &={N \over {48 b_0^2}}  \ ,\cr }
}
where
\eqn\schwarz{
D^{S}_{\sigma} (x)= {{\sigma^{\prime \prime \prime}} \over
{\sigma^{\prime}}} -{3\over 2}\biggl( {{ \sigma^{\prime \prime}} \over
{\sigma^{\prime} }}\biggr)^2
}
is the Schwarzian derivative and $\sigma^{\prime}$ denotes
$\partial \sigma / \partial x$.
Thus the limiting form is
unstable with a finite outgoing flux.
This  strongly suggests that
a naked singularity will form. In the following section
this will be confirmed by direct numerical calculation for
the shockwave geometry.

\newsec{Black Hole Evaporation }

In this section the equations \motion\ will be solved numerically
for incoming matter shock waves which form black holes, and then
subsequently evaporate. The same algorithm is used as in Section 3.

Now the boundary conditions are that $\rho$ and $\phi$ match onto
the vacuum \vac\ along the null line $\xp = 1$. Again at past null
infinity (or more practically large negative $\xm$)
the boundary condition is that $\dpl \phi$ and $\dpl \rho$
correspond to a classical black hole for $\xp > 1$.
Translated into conformal gauge, this condition is
\eqn\boundary{
\dpl\phi \sim -{1\over {\xp-\xm}} + {{4M}\over {(\xp-\xm)^2}}\biggl(
1- \log(\half(
\xp-\xm))\biggr) , \qquad \dpl\rho \sim {{2M}\over {(\xp-\xm)^2}} \ .
}
The boundary
condition is imposed in a region where the quantum effects
are always negligible so one may be sure
there is no extra incoming energy flux other than the
finite energy of the initial shock
wave.

{}From the usual adiabatic analysis the lifetime of these black holes
$t \sim M^3$. This is observed in these numerical studies as a
strong dependence of the lifetime on the mass, in contrast to
the CGHS case where $t\sim M$.
Since one is interested in the endpoint of the evaporation this
effectively restricts
one to the study of initial masses of order the Planck mass.

In
\fig{\fnine}{Paths of the singularity and the apparent horizon
for the model obtained from the 4d Einstein equations, with
$M=3$ and $N=2400$. The apparent horizon recedes and eventually
crosses the singularity.}
the paths of the singularity and the apparent
horizon $\dpl \phi =0$ are displayed.
The set of parameters chosen ($M=3$ and $N=2400$)
are believed to represent the
generic case.
These collide after a finite proper time
signaling the formation of a naked singularity.
The path of the singularity is represented by a line of large
constant $R$. The true line of
singularity lies at slightly larger $\xm$, and becomes tangent to a line
of constant $\xm$ at the moment when the singularity becomes naked.

In
\fig{\ften}{Curvature on the apparent horizon for the model
obtained from the 4d Einstein equations, with $M=3$ and $N=2400$.}
the curvature
on the apparent horizon is shown, which
appears to become singular for finite $\xp$.
The curvature decreases along a line of constant
$\phi$ as the mass of the black hole is
radiated away, as shown in
\fig{\feleven}{Curvature along a line of constant $\phi$ for the model
obtained from the 4d Einstein equations, with $M=3$ and $N=2400$.} .

\newsec{Conclusions}

In this paper, black hole evaporation in the context of
two quite different semiclassical two dimensional dilaton
gravity models has been studied.
The second of these is obtained by the dimensional reduction
of the four dimensional Einstein equations so is expected
to closely reflect physics in four dimensions.
The formation of a naked singularity appears
to be a rather generic feature of these semiclassical calculations.
This is in agreement with the results of the exactly solved
modified models
\refs{\callan{--}\suss}.
However, this seems to disagree with recent results of Hawking and Stewart
\nref\hawkstew{S.W. Hawking and J.M. Stewart, ``Naked and
Thunderbolt singularities in black hole evaporation'',
hepth@xxx/9207105, July 1992.} %
\refs{\hawkstew}
who, in the context of the CGHS-type models,
claim the apparent horizon persists, and a thunderbolt singularity
forms. This kind of endpoint is similar to a stable
remnant, but the singularity
spreads out to infinity rather than remaining
in a bounded region. This disagreement may arise because
in \refs{\hawkstew} the boundary condition on a line of constant
$\xm$ is imposed in a region where quantum effects are likely to
be strong.
One might expect this would be equivalent to sending in some
non-zero energy flux over a long period of time balancing
the outgoing flux and preventing the naked singularity from
appearing.
In this paper the boundary condition is imposed
in a region where quantum effects are negligible so one
can be certain the matter shock wave carries only a finite total
energy.

In any case, the semiclassical approximation only holds
until the remaining mass of the black hole is of order the
Planck mass. One might hope that higher order quantum
corrections will tame the naked singularity and the
residual energy might then be released as some kind of
gamma ray burst
(a rather scaled down version of the big bang
naked singularity). Due to the energy scale involved it is
unlikely this could be detected unless it occurred in our
close proximity.
If one takes the emergence of a naked singularity
seriously it is natural to ask whether
naked singularities appear as
nonperturbative solutions of string theory.
If this is true it may make string theory unpredictive
and require the introduction of a density matrix description
as in
\refs{\hawk}.

\bigskip
\bigskip
\centerline{\bf Acknowledgements}

The author wishes to thank Igor Klebanov for encouragement
and helpful discussions, Christoph Holzhey for some
stimulating conversations and Herman Verlinde for a critical
reading of the manuscript.
This research was supported in part by DOE grant DE-AC02-76WRO3072,
NSF grant PHY-9157482, and James S. McDonnell Foundation grant No.
91-48.

\vfill\eject

\appendix{A}{Numerical Algorithm}

The nonlinear partial differential equations to be solved take the
form
\eqn\des{
\dpl \dm u = f(u, \dm u,\dpl u)
}
with boundary conditions imposed along the lines $\xp=\xp_0$ and
$\xm=\xm_0$. For fixed $\xp$ \des\ may be regarded as a first order
ordinary differential equation for $\dpl u$. A fourth order
Runge-Kutta routine with adaptive stepsize in the $\xm$
direction is used to compute $\dpl u$. The efficiency of this
method relies on $u$ being known for arbitrary $\xm$. This is
accomplished using a fourth order rational function
interpolation routine.
The additional error introduced by
this interpolation is more than compensated by the improved
efficiency of the adaptive stepsize method. The derivatives
$\dm u$ are computed using a second order backward difference
method.

Having obtained $\dpl u$ as a function of $\xm$ again a
fourth order Runge-Kutta method with adaptive stepsize is used to
evolve the solution a step in the $\xp$ direction. Repeating
this procedure allows to the solution to be evolved out to
large $\xp$.

\listrefs
\listfigs
\bye
\end